\documentclass[usenatbib]{mnras}
\usepackage[T1]{fontenc}
\usepackage{ae,aecompl}
\usepackage{bm}
\usepackage{amsmath,amssymb}
\usepackage{dcolumn}
\usepackage[dvips]{graphicx}
\usepackage{epsf}
\usepackage{color}
\usepackage{epsf}
\usepackage{epsfig}
\usepackage{graphicx}
\usepackage{longtable}
\usepackage{psfrag}
\usepackage{txfonts}
\usepackage[Symbol]{upgreek}       

\newcommand{\be}{\begin{equation}}
\newcommand{\ee}{\end{equation}}
\newcommand{\ba}{\begin{eqnarray}}
\newcommand{\ea}{\end{eqnarray}}

\title[Testing the {standard model of cosmology} with the SKA]{{Testing the {standard model of cosmology} with the SKA: \\ the cosmic radio dipole}}

\author[C. A. P. Bengaly et al.]
{
 \parbox{\textwidth}{
  C. A. P. Bengaly$^{1}$\thanks{E-mail: \texttt{carlosap87@gmail.com}},
  T. M. Siewert$^{2}$,
  D. J. Schwarz$^{2}$,
  R. Maartens$^{1,3}$
 }
 \vspace{0.4cm}\\
 \parbox{\textwidth}{
 $^{1}$Department of Physics \& Astronomy, University of the Western Cape, Cape Town 7535, South Africa\\
 $^{2}$Fakult\"{a}t f\"{u}r Physik, Universit\"{a}t Bielefeld, Postfach 100131, 33501 Bielefeld, Germany\\
 $^{3}$Institute of Cosmology \& Gravitation, University of Portsmouth, Portsmouth PO1 3FX, United Kingdom
 }
}

\begin{document}
\label{firstpage}
\pagerange{\pageref{firstpage}--\pageref{lastpage}}
\maketitle

\begin{abstract}
The dipole anisotropy seen in the {cosmic microwave background radiation} is interpreted as 
due to our peculiar motion. The Cosmological Principle implies that this cosmic dipole signal 
should also be present, with the same direction, in the large-scale distribution of matter. 
Measurement of the cosmic matter dipole constitutes a key test of the standard cosmological 
model.  Current measurements of this dipole are barely above the expected noise 
and unable to provide a robust test. Upcoming radio continuum surveys with the SKA should be 
able to detect the dipole at high signal to noise. We simulate number count maps for SKA survey specifications in Phases 1 and 2, including all 
relevant effects. Nonlinear effects from local large-scale structure contaminate the {cosmic 
(kinematic)} dipole signal, and we find that removal of radio sources at low redshift 
($z\lesssim 0.5$) leads to significantly improved constraints. We forecast that the SKA could determine the kinematic dipole direction in Galactic coordinates with an error of 
$(\Delta l,\Delta b)\sim(9^\circ,5^\circ)$ to $(8^\circ, 4^\circ)$, depending on the sensitivity. The predicted errors on the relative speed are $\sim 10\%$. These measurements would significantly reduce the present uncertainty on
the direction of the radio dipole, and thus enable the first critical test of consistency between the matter and CMB dipoles.
\end{abstract}


\section{Introduction}\label{intro}

The Cosmological Principle underlying the standard model of the Universe requires that the CMB 
and the matter distribution should be isotropic on large scales, after we remove the cosmic kinematic 
dipole due to the motion of the Solar System relative to the cosmic rest-frame. 
In particular, this implies that the matter distribution should have a kinematic dipole with the same 
direction as that of the CMB. This constitutes a critical test of the foundations of the standard
cosmological model \citep{Ellis1984, Schwarz2015}.

Since the CMB dipole establishes the cosmic rest-frame of the early Universe at 
last photon scattering and the observed distribution of matter is probing the late Universe 
(at redshifts of order unity), the prediction that the early and late cosmic rest-frames should coincide
is still awaiting accurate observational verification.

The CMB dipole was measured with high accuracy by the {\it Planck} 
collaboration~\citep{Akrami:2018}: 
\ba 
&&v = 369.82 \pm 0.11 \,\mbox{km/s} \nonumber\\
&& \mbox{towards} ~~ (264.021 \pm 0.011\,,\,48.253 \pm 0.005)^{\circ}\,, \label{cmbd}
\ea
where the direction is given in Galactic coordinates.

This dipole is expected to be dominated by the kinematic contribution, which is $O(10^{2})$ larger 
than the intrinsic fluctuations in the standard $\Lambda$CDM model. Since the cosmic variance 
of the dipole is very large, significant non-kinematic contributions remain possible and need to 
be tested by other means. Probing the dipole of the matter distribution in addition to that of the 
CMB will help to tighten constraints on putative non-kinematic contributions.

\begin{figure*}
\includegraphics[width=0.33\linewidth]{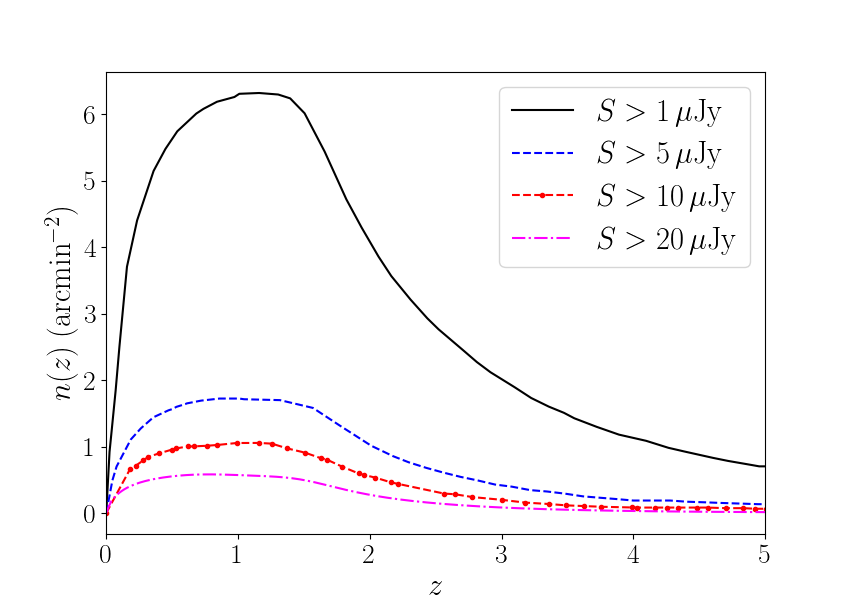}
\includegraphics[width=0.33\linewidth]{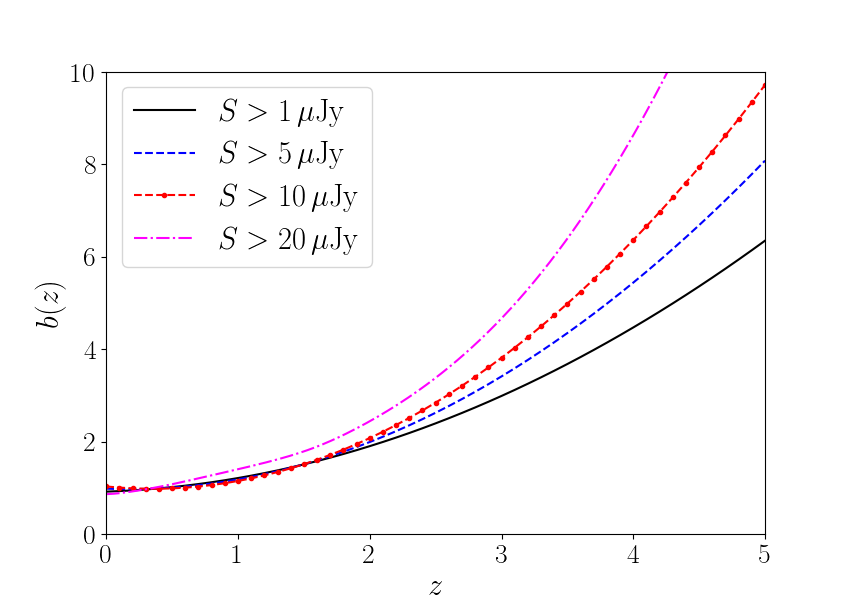}
\includegraphics[width=0.33\linewidth]{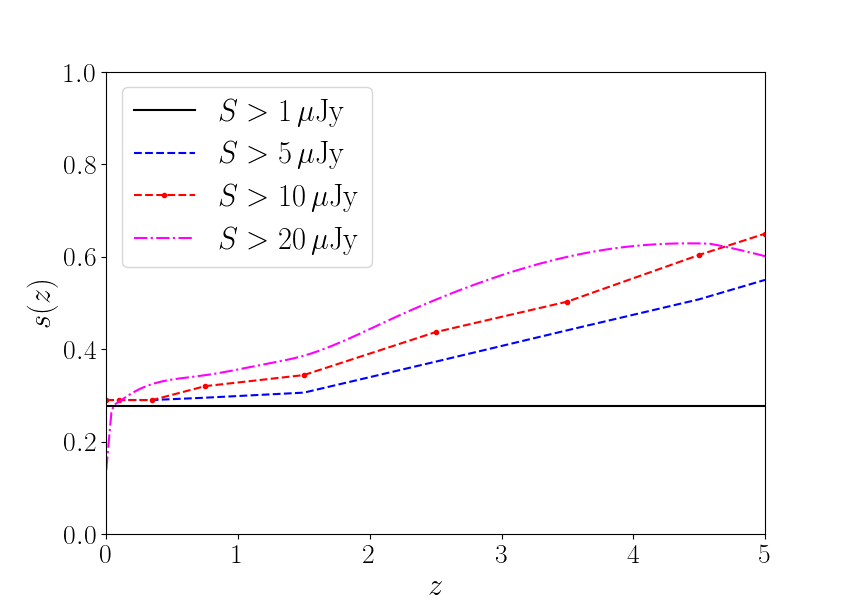}
\caption{Angular number density  (left), clustering bias  (centre), and magnification bias (right) as a function of redshift, for flux thresholds 1, 5, 10 and 20\,$\mu$Jy. }
\label{nbs}
\end{figure*}

The extragalactic radio sky offers an excellent opportunity to perform an independent test of the Cosmological Principle. 
The radio continuum dipole is expected to be dominated by the kinematic dipole. This is not the case for galaxy surveys at visible or infrared wavebands:  the number counts  in wide area surveys in those wavebands are dominated by objects at redshifts well below unity, so that the large-scale structure dominates over the kinematic signal. By contrast, radio continuum surveys have median redshifts above one, which suppresses the effect of local large-scale structure. Another advantage is that radio waves are not subject to extinction and thus the sky area that can be reliably observed by radio surveys exceeds that of optical and infrared surveys.

The largest available wide-area radio continuum surveys include the NRAO VLA Sky Survey (NVSS)~\citep{Condon1998} and the {TIFR} GMRT Sky Survey (TGSS)~\citep{Intema2016}.
Measurements performed using these and other radio continuum catalogues have found that the radio dipole is {compatible} with the CMB dipole direction, but the dipole amplitude is $2-5$ times larger than the signal observed in the CMB~\citep{Blake2002, Singal2011, Gibelyou2012, Rubart2013, Tiwari2014, Tiwari2015, Tiwari2016a, Colin2017, Bengaly2018b}. 
However, no significant evidence for anomalous anisotropy in galaxy counts was reported at lower redshift ranges, such as those probed by {visible} and infrared catalogues~\citep{Itoh2010, Gibelyou2012, Yoon2014, Alonso2015a, Yoon2015, Javanmardi2017, Bengaly2017, Bengaly2018a, Rameez2018}.

{Measurement of the continuum radio dipole is one of the {High-Priority Science Objectives} of the SKA,\footnote{{https://astronomers.skatelescope.org/wp-content/uploads/2016/12/SKA-TEL-SKO-0000122.pdf}} 
which can be extracted from the same type of survey that will allow us to measure or constrain 
primordial non-Gaussianity on the largest angular scales {\citep{Maartens2015}}. Here we 
focus on the prospects of measuring the dipole direction and amplitude by means of the SKA.}

We produce mock catalogues for SKA continuum surveys that include the effects of shot noise, large-scale structure and the kinematic dipole, and then we forecast the errors on SKA measurements of the radio dipole. This provides the details that underpin
the results that we presented in the SKA1 Cosmology Red Book~\citep{Bacon2018}.
{Here we include all the effects from large-scale structure, which were not included 
in the SKA Science Book. In addition,}
we extend the Red Book results to an alternative `optimistic' flux threshold for SKA1, as well as to `optimistic' and `realistic' flux thresholds for SKA2. 
This analysis {also} updates and extends previous forecasts presented in~\cite{Schwarz2015} 
(see also~\citealt{Crawford2009, Itoh2010, Yoon2015}).

\section{Analysis}\label{analysis}

\subsection{{Survey specifications}}

\begin{figure*}
\includegraphics[width=0.48\linewidth]{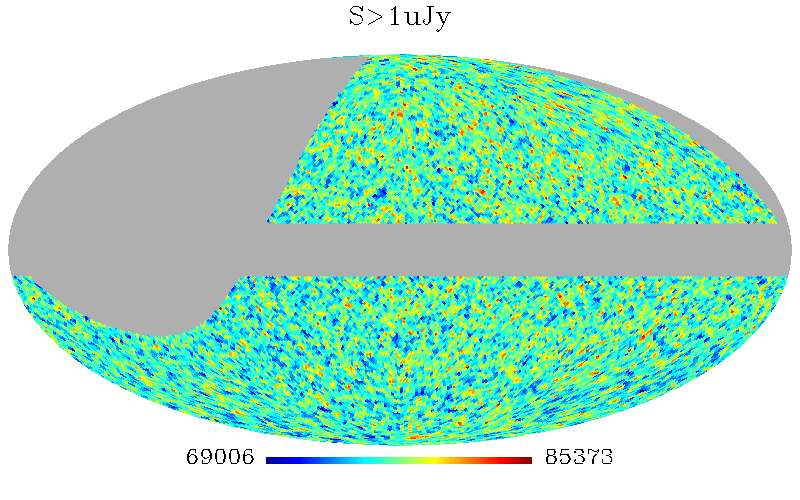}
\includegraphics[width=0.48\linewidth]{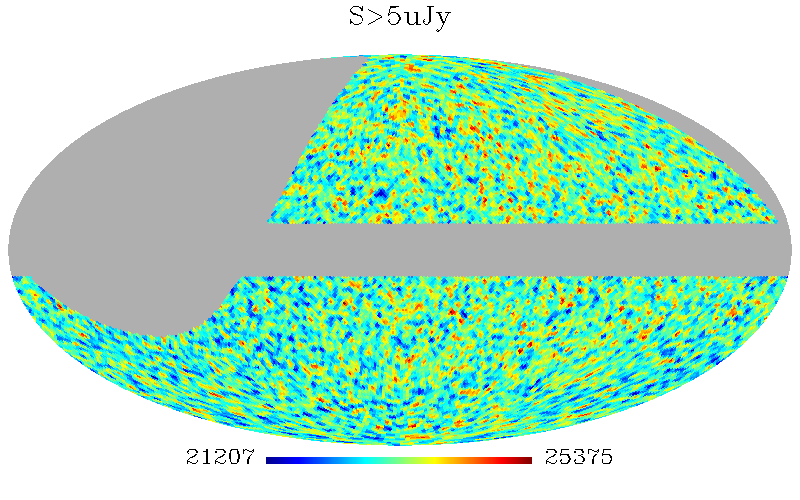}
\includegraphics[width=0.48\linewidth]{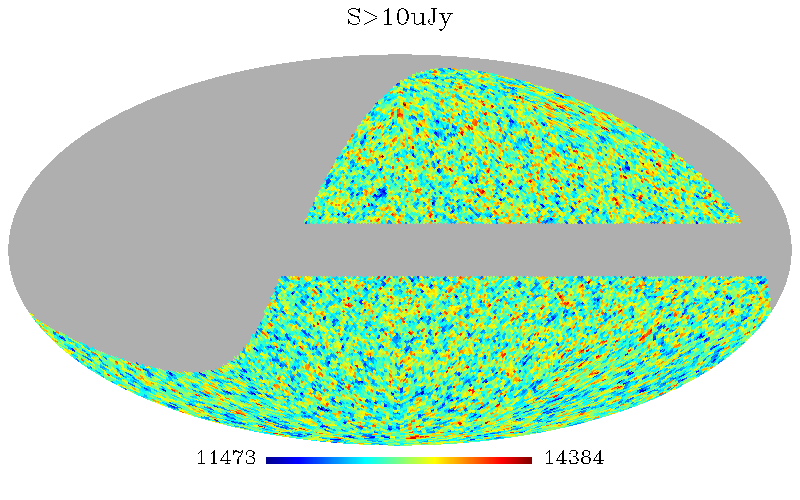}
\includegraphics[width=0.48\linewidth]{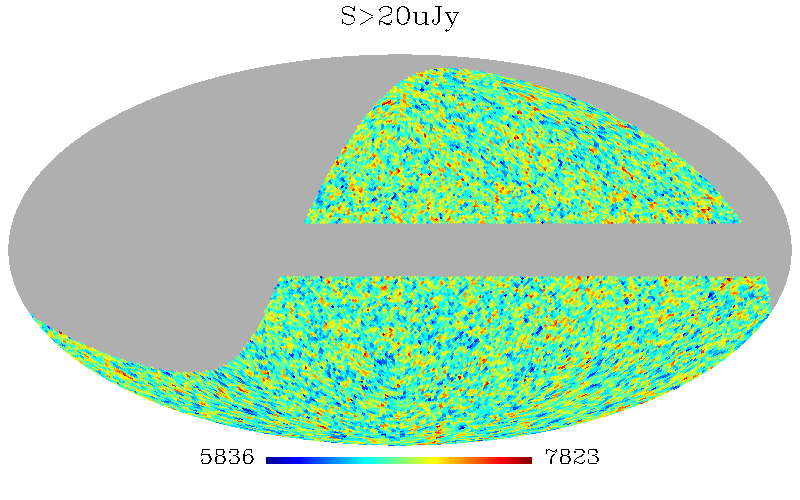}
\caption{Examples of SKA number count realisations for $S>1$ (upper left), $5$ (upper right), $10$ (lower left), and $20 \, \mu\mathrm{Jy}$ (lower right). }
\label{fig:SKA_maps}
\end{figure*}

Radio continuum surveys measure the integrated 
source flux in a frequency band, with rms noise
$\sigma$. For the proposed SKA1 survey~\citep{Bacon2018}:
\begin{eqnarray}
{\rm SKA1:}\quad && f_{\rm sky}\approx 0.63\,, \quad ~~ 350<\nu<1050\,{\rm MHz}\,,  \\
 && \sigma \sim 1\,\mu{\rm Jy/beam}\,,
 \quad 0< z\lesssim 5\,,
\end{eqnarray}
{where $f_{\rm sky}$ is the fraction of sky observed, $\nu$ is the observed radio frequency and $z$ is the redshift}.
{Specifications for SKA2 have not been formalised, but the rms noise is expected to be $\sim10$ times smaller and the sky coverage to reach $\sim 30,000\,$deg$^2$:}
\begin{eqnarray}
{{\rm SKA2:}\quad f_{\rm sky}\approx 0.75\,,  \quad \sigma\sim 0.1\,\mu{\rm Jy/beam}\,, \quad 0< z\lesssim 6\,.}
\end{eqnarray}
We assume that SKA2 has the same frequency range as SKA1. In this study, we assume that 
SKA1 (SKA2) will observe the sky up to $\rm{DEC}=15^{\circ}$ ($\rm{DEC}=30^{\circ}$), which give such $f_{sky}$ values, 
but we also mask low galactic latitudes ($|b| < 10^{\circ}$) in order to remove Milky Way objects. 
This amounts to effective sky area of $f_{\rm cut} \simeq 0.50 \; (0.63)$.

For the flux density threshold $S$, we adopt optimistic and realistic values as follows:
\begin{eqnarray}
{\rm SKA1:} && \mbox{optimistic:} ~  S>10\,\mu{\rm Jy}\,,~~ \mbox{realistic:} ~ S>20\,\mu {\rm Jy}, \label{smin}\\
{\rm SKA2:} && \mbox{optimistic:} ~  S>~1\,\mu{\rm Jy}\,,~~~ \mbox{realistic:} ~ S>~5\,\mu {\rm Jy}. 
\end{eqnarray}
In the SKA1 Red Book, $S>22.8\,\mu {\rm Jy}$ is used~\citep{Bacon2018}, which corresponds to the realistic case for SKA1.

\subsection{{Simulated data}}

In order to forecast the SKA dipole constraints, we generate 500 mock catalogues of radio number count maps. This is done in two steps:
\begin{itemize}
\item
Compute the theoretical angular power spectrum $C_\ell$ using 
{\sc CAMB sources}~\citep{Challinor2011}, which includes the effects on the 
observed number counts of redshift space distortions and lensing magnification.
\item
Input the theoretical $C_\ell$ and the redshift distribution of sources $n(z)$ to the lognormal 
code {\sc FLASK}~\citep{Xavier2016}, to generate mock SKA number count maps.
\end{itemize}

Computation of the theoretical $C_\ell$ requires as inputs the clustering bias $b(z)$, 
the redshift distribution of radio sources $n(z)$, and the magnification bias $s(z)$ 
(which determines the effect of lensing magnification on number counts). These quantities 
can be estimated by using the SKA Simulated Skies (S$^3$) data-base~\citep{Wilman2008}. 
We use the code\footnote{http://intensitymapping.physics.ox.ac.uk/codes.html} 
of~\cite{Alonso2015b}, which produces a semi-analytic fit to the S$^3$ luminosity function 
(details about the code are provided in their Appendix B). The results are shown in Fig.~\ref{nbs}.

For the fiducial cosmological model, we use the {\it Planck} 2015 best-fit flat 
$\Lambda$CDM parameters~\citep{Ade2016}. We compute $C_\ell$ in a single 
redshift bin spanning $0 < z < 5$ so that our window function $W(z)$ follows
\ba \label{eq:Wz}
W(z) &=& n(z)p(z), \\  
p(z) &=& 1, \quad \mathrm{for} \quad 0<z<5, \\
p(z) &=& 0, \quad \mathrm{for} \quad z>5. 
\ea
For each flux threshold $n(z)$ is shown in the left panel of Fig.~\ref{nbs}. We computed the $C_\ell$ in a single redshift bin, rather than splitting in different redshift bins as in~\cite{Bengaly2018b}, in order to better take the redshift-dependent clustering and magnification bias into account.

%
%

\subsection{Fiducial kinematic dipole}

\begin{figure*}
\includegraphics[width=0.48\linewidth]{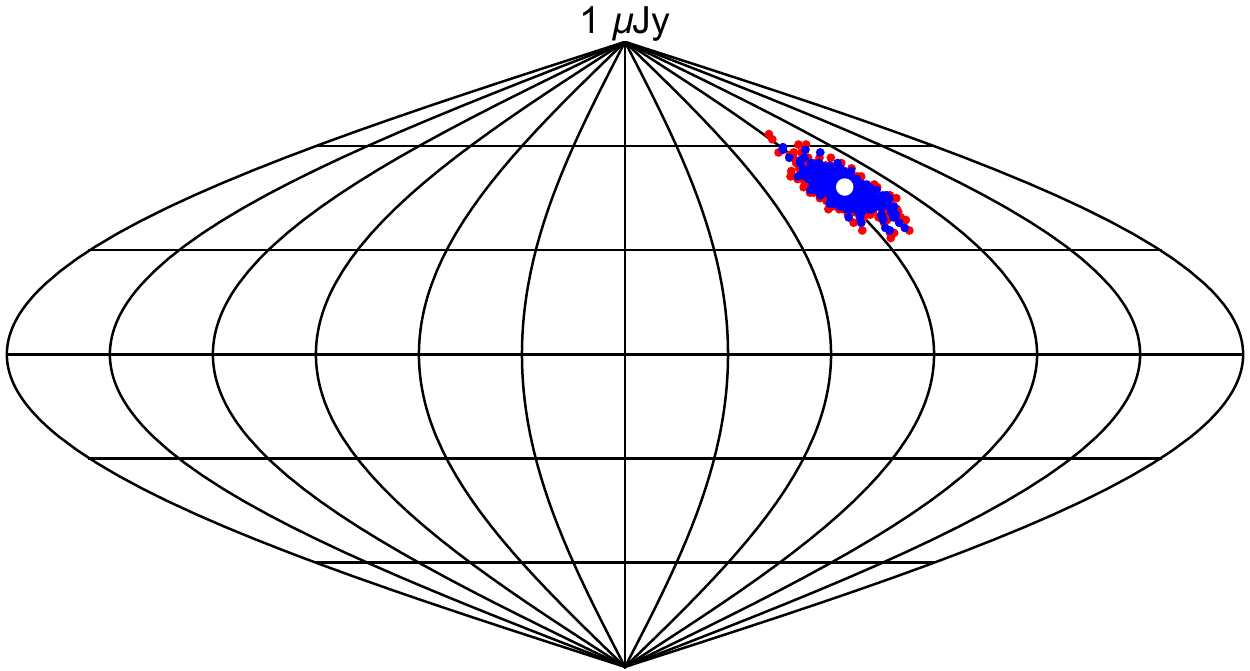}
\includegraphics[width=0.48\linewidth]{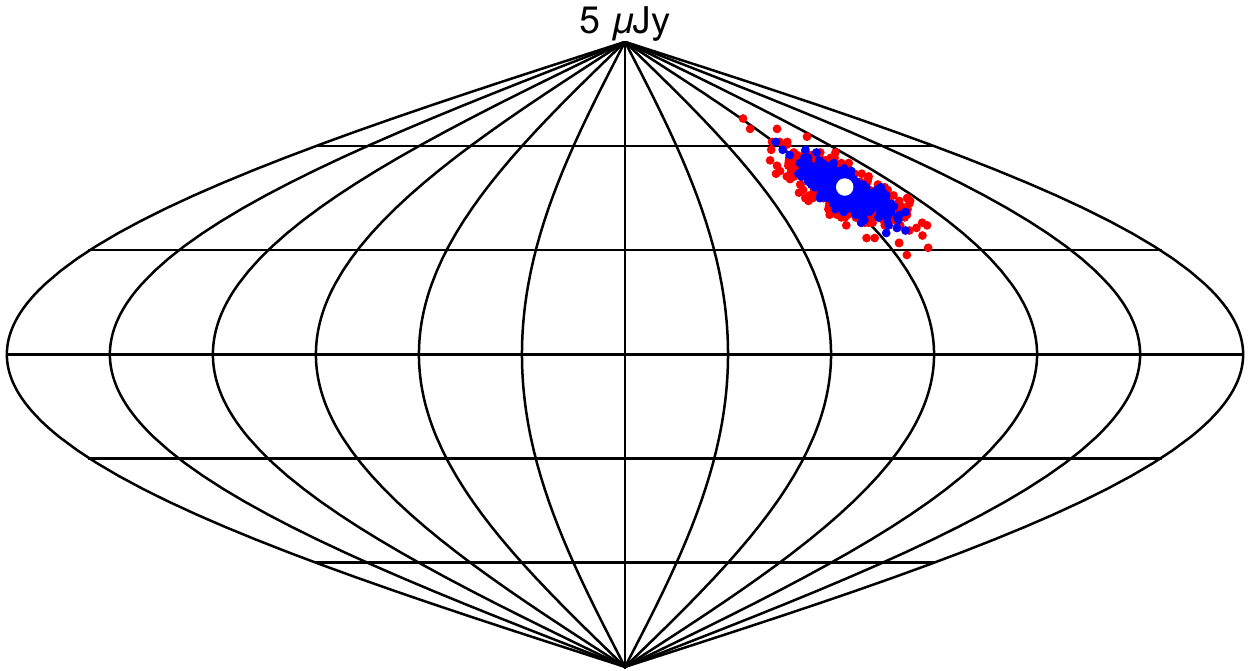}
\includegraphics[width=0.48\linewidth]{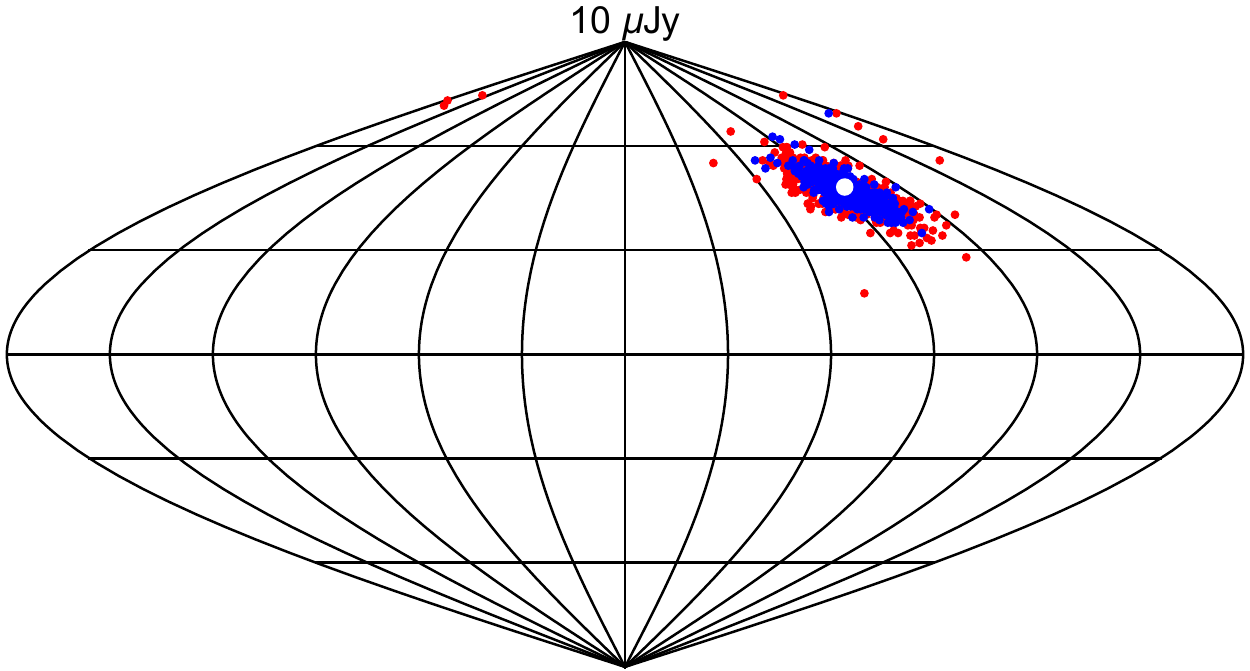}
\includegraphics[width=0.48\linewidth]{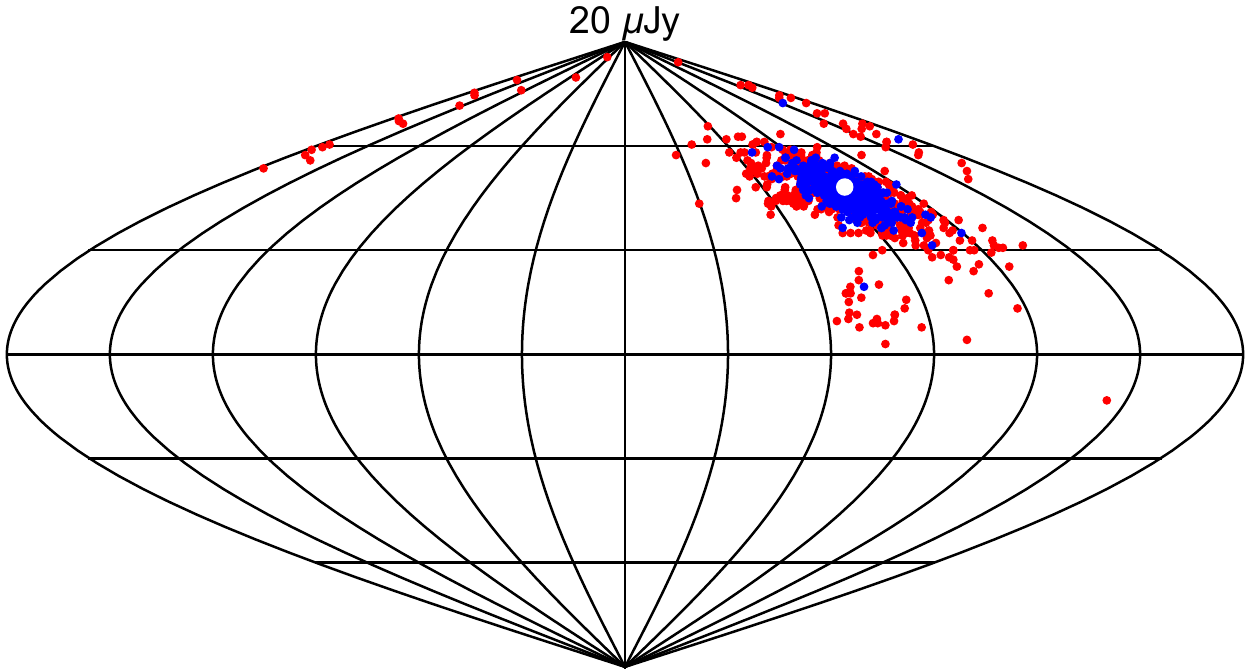}
\caption{Dipole directions from \eqref{dip_est_dom} for the flux limits indicated, based on 500 simulations each, in Galactic coordinates and stereographic projection.  Dots show the CMB dipole (white), and the kinematic dipole with (red) and without (blue) local structure.}
\label{fig:dipole_directions_TS}
\end{figure*}

\begin{figure*}
\includegraphics[width=0.48\linewidth]{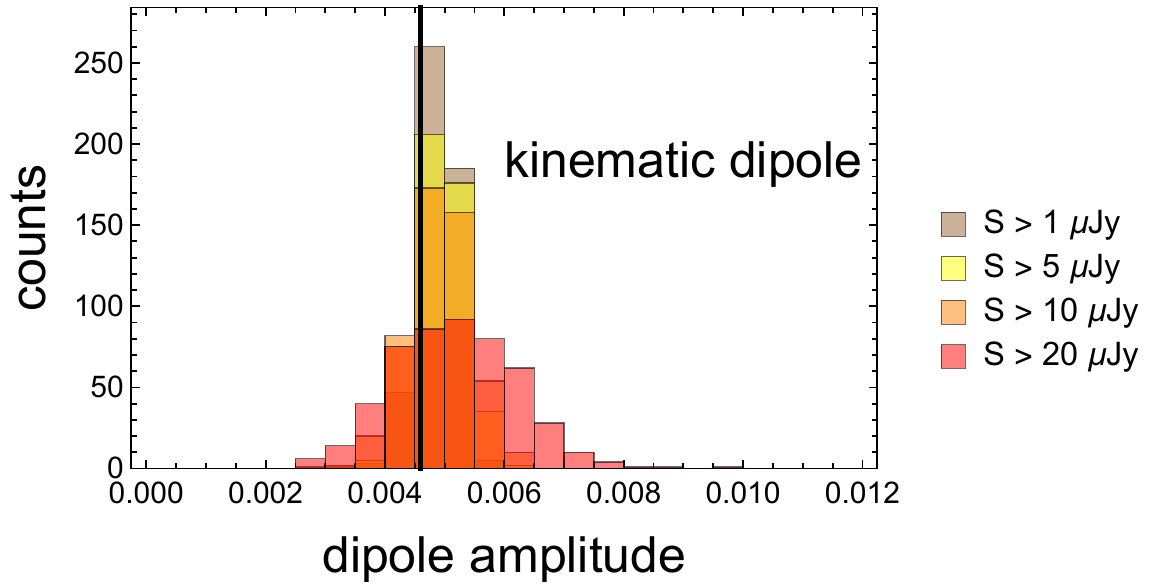}
\includegraphics[width=0.48\linewidth]{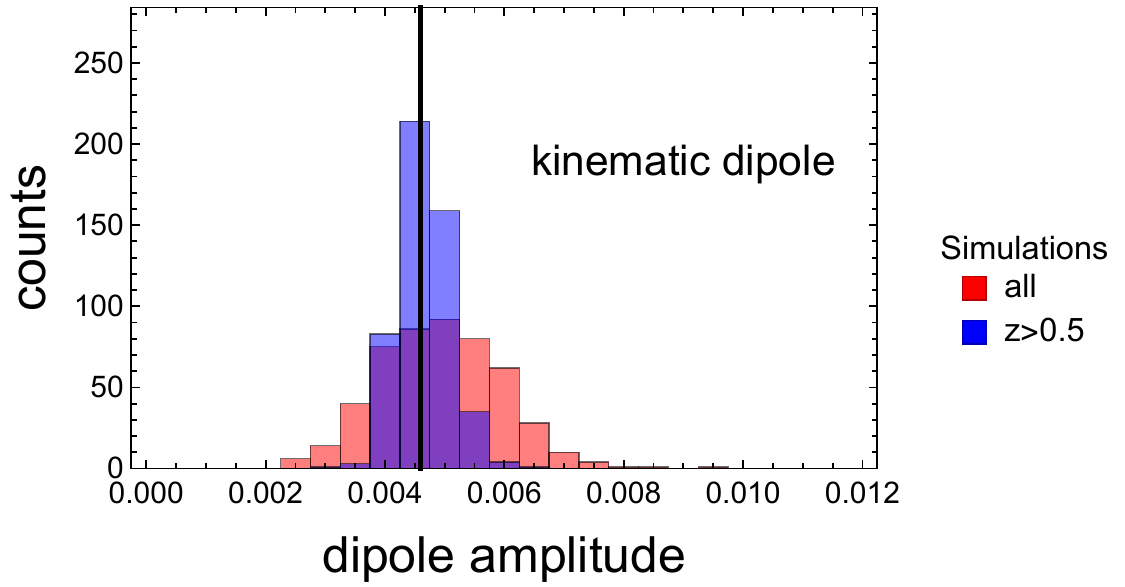}
\caption{Dipole amplitude corresponding to Fig.~\ref{fig:dipole_directions_TS}. {\em Left:} Histogram for 4 flux thresholds, including all sources ($z>0$). The fiducial value \eqref{akin} is the black line. {\em Right:} $S>20\,\mu$Jy case, with and without $z<0.5$ sources.}
\label{fig:dipole_amplitudes_TS}
\end{figure*}

Following~\cite{Bengaly2018b}, the effect of the kinematic dipole is input in the mock maps through a 
dipole modulation of the number counts. Imposing a dipole modulation in pixelised source 
counts, or individually boosting each source according to the fiducial dipole signal, lead to 
negligible differences in the dipole estimates. We chose the former for computational 
efficiency reasons. Details of the pixel modulation and a comparison to applying a boost to 
each source are provided in Appendix A

To first order in perturbations, the modified source counts due to a boost of the observer are 
\be \label{nobs}
N_{\mathrm{obs}}(\bm{n},>\!S) = N_{\mathrm{rest}}(\bm{n},>\!S)\left[1 + A\,\bm{n} \cdot \hat{\bm{\beta}}\right]\,, 
\ee
where $N$ is the source count per solid angle {about the unit direction
 $\bm{n}$} and above a flux threshold $S$, and $\bm{v} =\bm{\beta}c$ is the velocity of the Solar System observer relative to the CMB rest-frame. The boost amplitude is~\citep{Ellis1984}:
\be
\label{amp}
A = \big[2 + x(1+\alpha)\big]\, \beta \,.
\ee
The number count contrast is defined by 
\be
\delta(\bm{n},>\!S) = {N(\bm{n},>\!S)-{\bar{N}(>\!S)} \over {\bar{N}(>\!S)}}\,,
\ee
where
$\bar{N}$ is the average number of sources per solid angle. Then \eqref{nobs} gives
\begin{eqnarray}
\label{delo}
\delta_{\mathrm{obs}}(\bm{n},>\!S) = \delta_{\mathrm{rest}}(\bm{n},>\!S) + A\cos{\theta}\, 
,~~ \cos\theta =\bm{n} \cdot \hat{\bm{\beta}}\,.
\end{eqnarray}

The expression \eqref{amp} assumes that the flux density and counts are given by power laws:
\begin{eqnarray}
S \propto \nu^{-\alpha}\,,\quad \bar{N}(>S) \propto S^{-x}\,.
\end{eqnarray}
{The fiducial radio kinematic dipole amplitude is computed from \eqref{amp}  using the SKA1 Red Book values~\citep{Bacon2018},
\begin{equation}\label{xal}
x=1\,,\quad \alpha=0.76\,,
\end{equation}
for all flux thresholds, while $\beta$ and $\hat{\bm{\beta}}$  are taken from the CMB dipole values in \eqref{cmbd}. Then the fiducial dipole amplitude is 
\ba\label{akin}
A_{\rm kin} = {4.62} \times 10^{-3}\,.
\ea
 
At low redshifts, the kinematic dipole is dominated by the dipole induced by the nonlinear influence of local large-scale structure. The local structure dipole is a contaminant to the cosmic kinematic signal, whose effects decay with redshift. We found that an effective way to suppress this contamination is to excise all sources at small redshift $z< z_\mathrm{cut}$. Here we present results mainly for $z_\mathrm{cut} = 0.5$, but we have also tested values of $0.1, 0.2$, and $0.3$. 
This is done by performing the simulation procedure described in section 2.2 for $z_\mathrm{cut}<z<5.0$. The consequent increase in shot noise 
is not significant: unlike current surveys, SKA surveys are not dominated by shot noise until 
$\ell>500$ (see Fig.~2 in~\citealt{Pant2018}).  
The removal of low-$z$ sources should be feasible by cross-correlating radio continuum data with optical, infra-red or future 21cm data, as discussed in the SKA1 Red Book~\citep{Bacon2018}. In~\cite{Duncan18}, it is shown that such cross-correlation allows a good estimate of radio source photo-$z$, specially at $z<0.5$.

\begin{table*}
    \centering
    \caption{Averaged dipole direction (Galactic coordinates) and amplitude for 500 simulations at 4 flux thresholds, with and without $z<0.5$ sources. The average number of total sources, $N_{\rm tot}$, at each flux threshold, and for each sample, is also given.}
    \begin{tabular}{ccccccccccc}
        \hline\hline
        Sample & & $N_{\rm tot}$ & & ${S >}$ & & $l$ & & $b$ & & $A$ \\
	& & $(10^{9})$ & & ($\mu$Jy) & & (deg) & & (deg) & & $(10^{-3})$ \\\hline \hline 
	&&&&&&&&\\
        full & & $2.37$ & & $1.0$ & & $264.87 \pm 6.47$ & & $47.42 \pm 4.33$ & & $4.64 \pm 0.33$\\
        $z\geq0.5$& & $2.07$ & & $1.0$ & & $264.64 \pm 5.57$ & & $47.37 \pm 3.67$ & & $4.64 \pm 0.30$\\\hline
        full & & $0.72$ & & $5.0$ & & $265.15 \pm 8.43$ & & $47.29 \pm 5.58$ & & $4.67 \pm 0.44$\\
        $z\geq0.5$& & $0.62$ & & $5.0$ & & $264.84 \pm 5.77$ & & $47.43 \pm 3.85$ & & $4.64 \pm 0.30$\\\hline
        full & & $0.33$ & & $10.0$ & & $264.50 \pm 12.71$ & & $47.08 \pm 6.18$ & & $4.66 \pm 0.55$\\
        $z\geq0.5$& & $0.29$ & & $10.0$ & & $264.56 \pm 7.34$ & & $47.20 \pm 3.97$ & & $4.62 \pm 0.38$\\\hline
        full & & $0.18$ & & $20.0$ & & $263.86 \pm 25.08$ & & $45.50 \pm 12.89$ & & $4.93 \pm 1.03$\\
        $z\geq0.5$& & $0.15$ & & $20.0$ & & $265.49 \pm 8.65$ & & $46.83 \pm 4.64$ & & $4.65 \pm 0.45$\\\hline
        &&&&&&&&\\
       {fiducial}  & & $-$ & & & & $264.02$ & & $48.25$ &&  $4.62$ \\\hline
    \end{tabular}
    \label{tab:dipole_estimates_TS}
\end{table*}

\subsection{Estimator}

Linear estimators of the dipole suffer from bias, as 
discussed in~\cite{Rubart2013}. Quadratic estimators that are free from
this problem can be constructed. A straightforward choice of such an
estimator for the dipole in pixel space is to vary the monopole and dipole of number counts so as to find
\be\label{dip_est_dom}
\min \sum_p \frac{ \left[ N_{p}(\bm{n},>\!S) - \bar{N}(>\!S)\,\left(1 + A\cos{\theta_{p}}\right) \right] ^2}
{\bar{N}(>\!S)\,\left(1 + A\cos{\theta_{p}}\right)}\,. 
\ee
The sum is taken over all unmasked pixels and the monopole $\bar{N}(>\!S)$  is an {estimate of the all-sky} average.

Comparing \eqref{dip_est_dom} with \eqref{nobs}, we have approximated 
$N_{\mathrm{rest}}(\bm{n},>\!S)$ by $\bar{N}(>\!S)$: in other words,
the expression \eqref{dip_est_dom} neglects clustering and effectively  assumes  a Poisson
distribution. In principle one could model deviations from such a Poisson distribution due to 
large-scale structure by  including the full covariance matrix, using the theoretically expected 
angular two-point correlation function.\footnote{Note that we do not assign any physical meaning 
to the minimum of the estimator in this work, i.e.,~we are not calculating p-values.} We leave 
the inclusion of clustering for a future study.

Instead, here we suppress the effect of large-scale structure on the estimator by using rather large 
pixels and downgrading each simulated map to $N_{\rm side} = 16$ for computational efficiency.
Then we minimise the estimator {for each simulation on a three-dimensional grid of dimension $49152 \times 6 \times 20$. The dipole direction is probed along the 
pixel centres defined for $N_{\rm side} = 64$. The mean distance between neighbouring pixel centres at this resolution 
is $0.92^\circ$. Using the {\sc HEALPix} `ring' scheme, we effectively reduce the two angular dimensions 
to a one-dimensional index that probes the full sphere. The other two grid directions are the monopole and dipole amplitudes, 
which vary for the dipole amplitude in steps of $5\times10^{-4}$.}

\section{Results}\label{results}

The results of 500 simulations for each flux threshold are shown in Fig. \ref{fig:dipole_directions_TS} for the dipole direction and Fig.~\ref{fig:dipole_amplitudes_TS} for the dipole amplitude. The numerical values are summarised in Table~\ref{tab:dipole_estimates_TS}. 

The central {values and spread} shown are calculated via the arithmetic mean and 
standard deviation from the simulations for each sample, which are straightforward for the latitude 
$b$ and amplitude $A$. 
For the longitude $l$, we need to account for the fact that close to the pole, a small 
shift can lead to a large difference in $l$, whereas close to the equator 
$l$ is a good measure.
We therefore use a weighted mean and standard deviation, 
defined as follows:
\ba
\bar{l}&=& \frac{1}{k}\sum_{i}w_i l_i\,,\quad w_i = \frac{k\,\cos b_i}{\sum_{i}\cos b_i}\,, \\
\sigma_l^2 &=& {\frac{1}{k-1}\sum_{i}w_i\left(l_i-\bar{l}\,\right)^2}\,.
\ea 
Here the sums are over $i=1,\cdots,k$, where $k \,(=500)$ is the number of realisations.}

The  dipole amplitudes are consistent among the four flux thresholds, but increase slightly with increasing flux limits. Our results recover the assumption of a  kinematic radio dipole with expected amplitude given by \eqref{akin}. We find the best agreement for the lowest flux threshold and without local structure:
\be
A= (4.64\pm0.30)\times10^{-3}, \quad \mbox{for}~S>1\,\mu{\rm Jy}~\mbox{and}~z\geq 0.5\, .
\ee 
The contribution from local structure increases with the flux limit, becoming significant for the highest flux limit. 

\begin{figure}
\includegraphics[width=\linewidth]{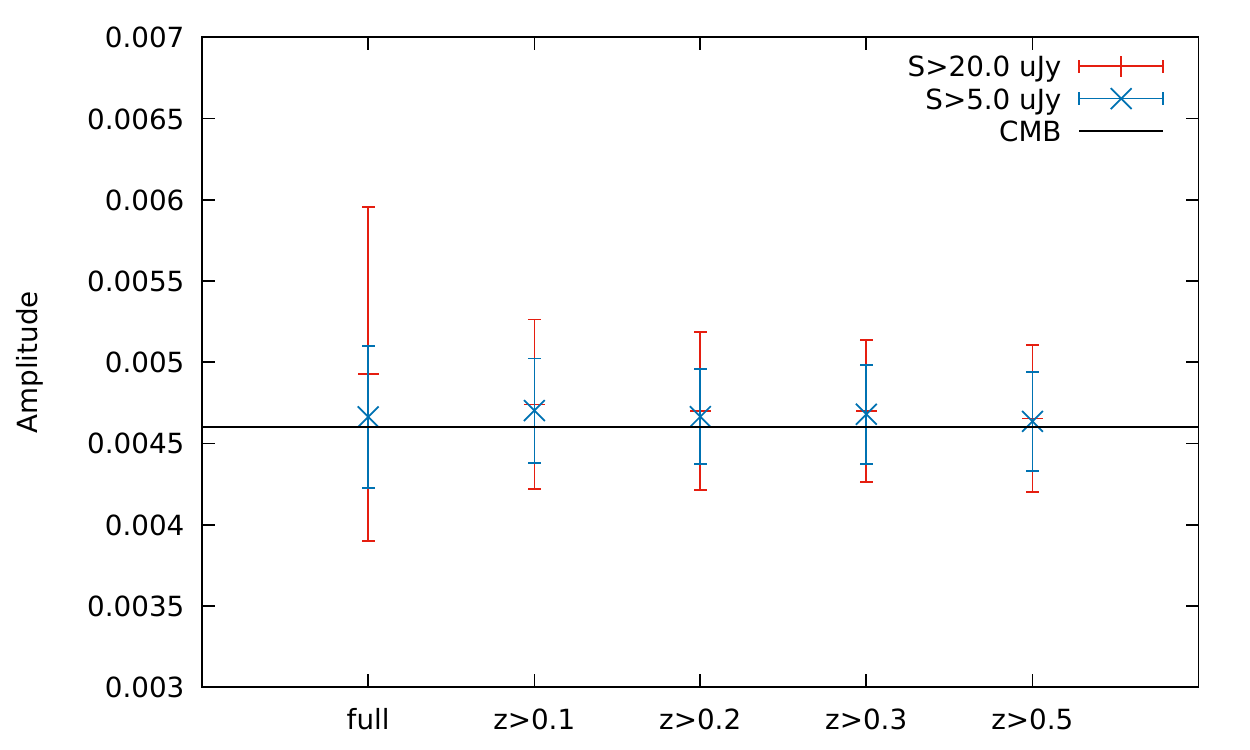}
\caption{Constraints on the kinematic dipole amplitude as a function of lower cut in redshift for 
the realistic SKA1 and SKA2 simulations. The results become more stable from $z_\mathrm{cut} > 0.1$ onward 
since the most strongly clustered structures are eliminated at this redshift range.}
\label{fig:cuts}
\end{figure}

The recovered dipole directions are also in agreement with the CMB dipole position, given in \eqref{cmbd}, for all flux thresholds. For the lowest flux sample, and without local structure, 
we measure the dipole direction {with degree-level accuracy}:
\ba
 && (l,b)=(264.64 \pm 5.57\,, \, 47.37 \pm 3.67^\circ),  \nonumber\\
&& \mbox{for}~S>1\,\mu{\rm Jy}~\mbox{and}~z\geq 0.5\, .
\ea 
When we remove local structure, the error on the dipole position drops significantly and the 
directions become more consistent with the CMB dipole direction. As is evident in Fig. 
\ref{fig:dipole_directions_TS}, the spread of the individual dipole directions is larger for higher flux 
limits. The simulations without local structure (blue) are still concentrated around the CMB dipole 
direction (white), while the full simulations (red) begin to spread over the northern galactic hemisphere. 
We also found that most of the dipole estimate improvement occurs when we apply 
$z_\mathrm{cut}=0.1$. This is expected since the clustering of sources becomes stronger at the 
lowest redshift ranges,as previously investigated in~\cite{Blake2002, Yoon2015, Tiwari2016a}. 
Our choice of $z_{\rm cut}=0.5$ is on the optimistic side as we implicitly 
assume that we can get photo-z redshifts for all sources. 
Figure \ref{fig:cuts} shows how the accuracy of the 
reconstruction of the kinematic dipole amplitude changes with the choice of $z_\mathrm{cut}$ for the 
$20$ and $5 \mu$Jy simulations. 
This result agrees with~\cite{Tiwari2016a}, in which most of the source clustering contamination occurs at $z \leq 0.1$ 
- but note that the NVSS sample is comprised of AGNs, whereas the radio sky should be dominated by star-forming galaxies at the SKA flux thresholds.

In order to have a comparison with the estimator \eqref{dip_est_dom}, we present forecasts 
using an alternative estimator in Appendix B.

We have checked the dependence of the results on the choice of the grid
steps in monopole and dipole amplitude. Refining the grid led to small
changes both in central value and spread, but
even grid refinement by a factor of $5$ in the case of the dipole amplitude
produced results that are in good agreement with those presented. As
such a refinement also leads to an corresponding increase in
compute time, there is a trade off between the number of simulations and
the accuracy of the dipole estimation.

\section{Conclusions}\label{sec:conc}

The standard model of cosmology is highly successful in accounting for current observations of the CMB and large-scale structure. Nevertheless, it remains of crucial importance to also test the theoretical consistency of the model, using current and upcoming observations. A fundamental test  is to probe the consistency between the kinematic dipoles measured in the CMB and in the large-scale structure. A mismatch between these dipoles could indicate either a violation of the Cosmological Principle on the large scales where it is required to hold, or a sign of new physical features in the primordial Universe.

The CMB kinematic dipole has been measured with exquisite accuracy by {\it Planck}. Measurement of this dipole in galaxy surveys faces formidable problems. The survey needs to cover a significant fraction of the sky with low shot noise, in order to make a detection of the dipole. In addition, the survey must reach redshifts significantly above $z\sim 1$, in order to suppress the strong contamination of the kinematic dipole by local structure. 

These requirements are in principle met by wide radio continuum surveys. The existing all-sky NVSS and TGSS are unable to deliver the signal-to-noise needed for a robust measurement of the kinematic dipole direction and amplitude. We have investigated the capacity of next-generation SKA surveys to make a measurement that allows for a consistency test. By simulating SKA number count maps that include clustering, the fiducial kinematic dipole and shot noise, we have shown that SKA1 should be able to make a measurement with $\sim 10\%$ accuracy on the amplitude and $(\Delta l,\Delta b)\sim(9^\circ,5^\circ)$ to $(8^\circ, 4^\circ)$ accuracy on the direction. Although this is well behind  {\it Planck} precision, checking 
the consistency of the two directions within degree level would be a remarkable test. SKA2 delivers increased precision, but not by a significant margin: in other words, this critical consistency test will already be feasible in Phase 1 and does not have to wait for Phase 2.

A key aspect of our constraints is the significant improvement that we achieved by excising sources at 
$z< 0.1$ to $0.5$ in order to suppress the contamination from local structure. 
Cross-correlating the SKA map with current and upcoming optical/infrared and low-$z$ 21cm 
data should allow for this excision to be implemented effectively. 
As shown in~\cite{Duncan18}, multi-wavelength identification of low-$z$ 
radio sources can be done with good precision and low-$z$ radio objects are very unlikely to be
classified as a high-$z$ ones. Hence, one should be able to correctly identify a large fraction 
of the low-$z$ objects. Note also that this is a conservative estimate based on 
LOFAR observations, hence this procedure should be even more efficient in the higher 
frequency range that SKA will observe. The effect of partial local 
structure suppression on the cosmic dipole will be further investigated in the future.

We did not model any systematic instrumental effects here; some of them are addressed in the SKA1 
Science Red Book and are shown to not be show-stoppers. A first assessment of some of these 
effects, such as flux calibration errors, was performed in~\cite{Schwarz2015}. 
It was found that flux calibration errors lead to higher dipole amplitude, as also explored in~\cite{Bengaly2018b}.
We plan to revisit this analysis once observed data from MIGHTEE~\citep{Jarvis17} become available, 
since it will better inform us about the actual SKA flux calibration uncertainty. 
A future study should test if there is any unexpected cross-talk between possible systematics and large-scale structure that could 
undermine accurate measurement of the kinematic dipole.


In order to include all instrumental and observational effects in radio 
dipole and SKA radio source catalogue forecasts,
it will be necessary to simulate the fluctuations in position and flux 
calibration, limitations due to dynamic range in the
vicinity of bright sources, foreground emission, scan strategy and 
source extraction. However, these will need to start from an
intermediate level in the data analysis, e.g.  from calibrated 
maps for individual pointings.
It will not be possible to do full end-to-end simulations for a full SKA 
survey in a similar way as full end-to-end simulations are done for CMB 
experiments, given the huge raw data rates of SKA dishes.
We shall revisit the dipole forecasts from simulations that incorporate such features in the future.


\subsection*{Acknowledgements}
We thank Song Chen for comments, and Mario Santos, Mario Ballardini, and Nidhi Pant for very useful discussions. CB and RM acknowledge support from the South African SKA Project and the National Research Foundation of South Africa (Grant No. 75415). RM was also supported by the UK Science \& Technology Facilities Council (Grant No. ST/N000668/1). TS and DJS gratefully acknowledge support from the Deutsche Forschungsgemeinschaft (DFG) within the Research Training Group 1620 ``Models of Gravity''.

\newpage
\appendix 

\section{Source vs. pixel boost}

\begin{table*}
	\centering
	\caption{Dipole estimates for boosted source positions and boosted pixel count maps based on 
	100 random cataloges with $N = 10^6$. $\bar N(>S)$ denotes the mean number of sources 
	reconstructed after the boost. (r) and (nr) indicate that counts in cells have been rounded to 
	the next integer or not rounded after the pixel boost.}
		\begin{tabular}{lcccc}
		\hline\hline
		boost & $\bar N(>S)$ & $l$ & $b$ & $A$ \\
		method            &  & (deg) & (deg) & $(10^{-3})$ \\\hline\hline
		source & $899985. \pm 283. $ & $260.87 \pm 45.08$ & $45.05 \pm 17.81$ & $5.70 \pm 1.70$\\
		pixel (r)    & $899977. \pm 287.$ & $260.79 \pm 44.75$ & $44.75 \pm 18.25$ & $5.50 \pm 1.70$ \\ 
		pixel (nr)  & $899978. \pm 288.$ & $260.86 \pm 45.05$ & $44.53 \pm 17.67$ & $5.70 \pm 1.70$ \\ \hline
		\hline
	\end{tabular}
\label{tab:dipole_estimates_boost}
\end{table*}

In order to save computational resources we include the kinematic dipole by 
boosting the pixel counts according to (\ref{nobs}) instead of boosting individual sources. 
The exact and approximate pipelines produce slightly different results.

The exact procedure would require the following steps: 1. produce a mock point source catalogue 
that contains positions and flux densities of all sources (it must contain also flux densities below the 
flux density threshold that is applied later on), 2. boost the fluxes and positions to the new observer rest frame, 3. apply a flux threshold, 4. produce a pixel map. 

The approximate procedure adopted in this work involves the following steps: 
1. take the same catalogue, 2. apply a flux density threshold, 3. produce a pixel map, 
4. apply the pixel boost according to equation (\ref{nobs}), and 
5. round to the next integer value in the pixel counts.

In order to compare source and pixel boost, we produced 100 isotropic 
random catalogues with a flux density distribution given by $N(>S) \propto 1/S$, i.e. $x= -1$, 
and a total of $N = 10^6$ sources. In order to end up with a large enough number of sources per 
pixel, such that the boost can have still an effect after rounding to an integer number, we perform 
the test for maps with $N_{\rm side} = 16$. (Note that $N > 10^8$ for all SKA simulations considered 
in this work.) The result is presented in table \ref{tab:dipole_estimates_boost}, where we 
also show the difference between rounding to integer values (r) or not rounding (nr).

The result is that all methods agree very well with each other and give very similar mean 
values and the variances. The variances observed in this test exceed those of the 
SKA simulations by a large amount, which is due to the much larger shot noise contribution, which 
is at least an order of magnitude smaller in all SKA simulations of this work. The 
reconstructed amplitude is also larger than the one reconstructed in the SKA simulations, 
which is also due to the smaller sample size. We see that an implementation of our simulation 
pipeline without rounding seems to produce a slightly better agreement between both procedures 
and slightly more significant dipole estimates. The statistical results are also confirmed when 
individual maps are compared. The maximal difference found in a pixel is of the order of the shot 
noise in all cases that we inspected.

We also compared the difference between source and pixel boosting for $N_{\rm side}=64$ grid resolution. We found fully consistent results with the $N_{\rm side} = 16$ case within errors. The rounding to integer number counts, however, could not be tested for $N_{\rm side}=64$ because the effect of the kinematic dipole in each individual cell 
changes the count by less than unity. This problem does not appear for the SKA simulations, as the 
number of counts per cell are so large that even a $10^{-3}$ effect exceeds unity in most cells (except 
normal to the dipole, of course).

We conclude that our method works very well for the purpose of simulating the kinematic dipole. We 
did not test its effect on higher multipoles that might also be of interest and leave that to a future study.

\section{Alternative estimator}

\begin{figure*}
\includegraphics[width=0.48\linewidth]{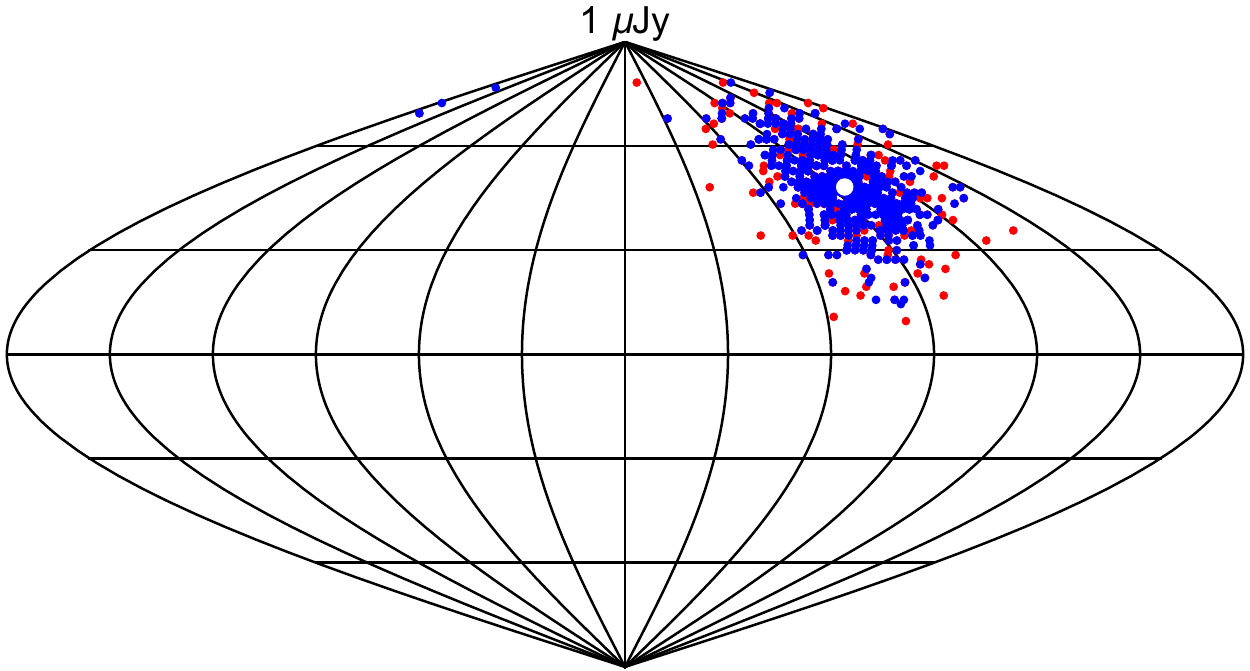}
\includegraphics[width=0.48\linewidth]{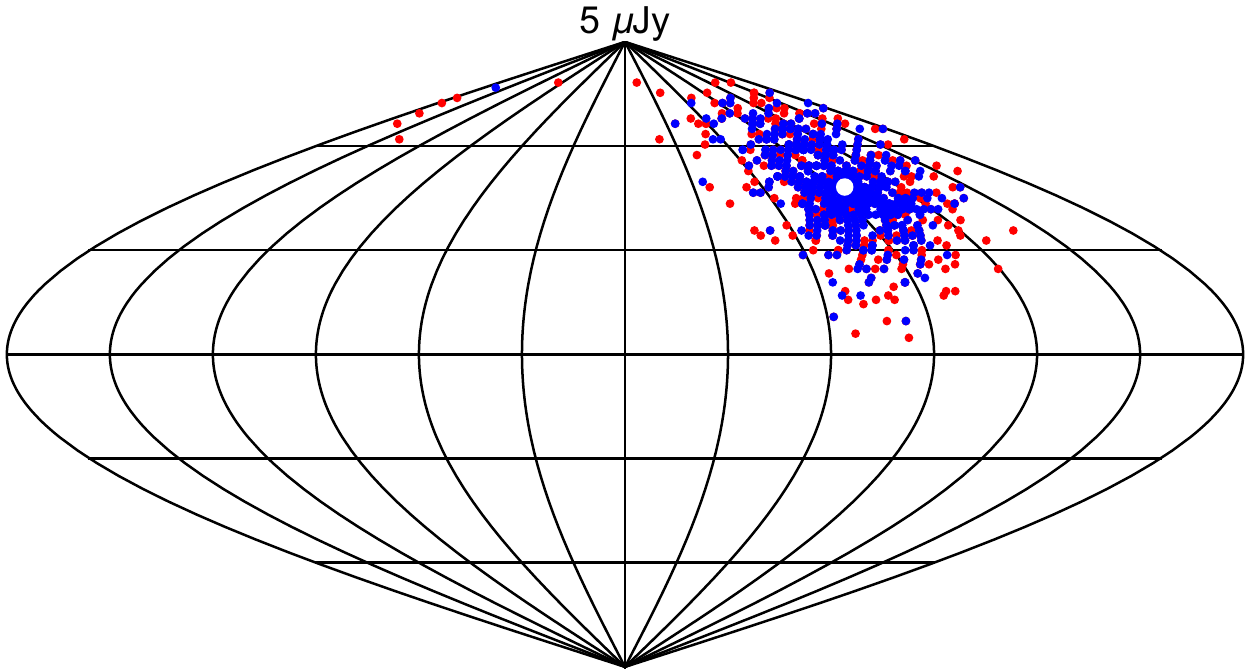}
\includegraphics[width=0.48\linewidth]{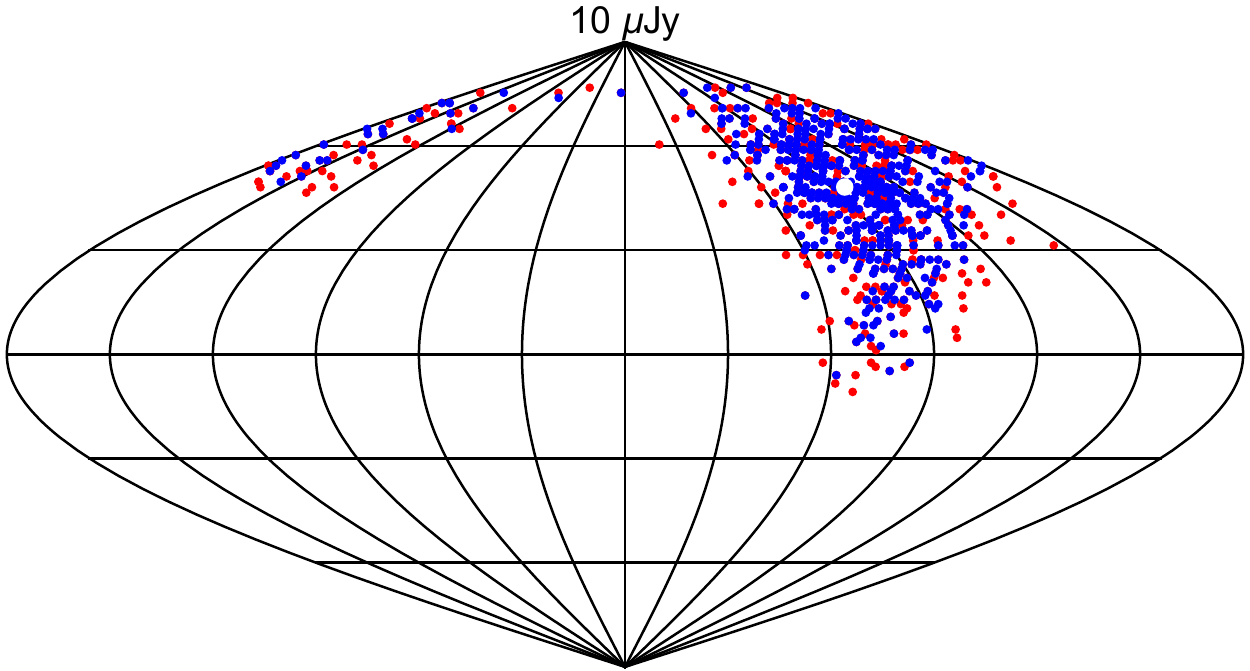}
\includegraphics[width=0.48\linewidth]{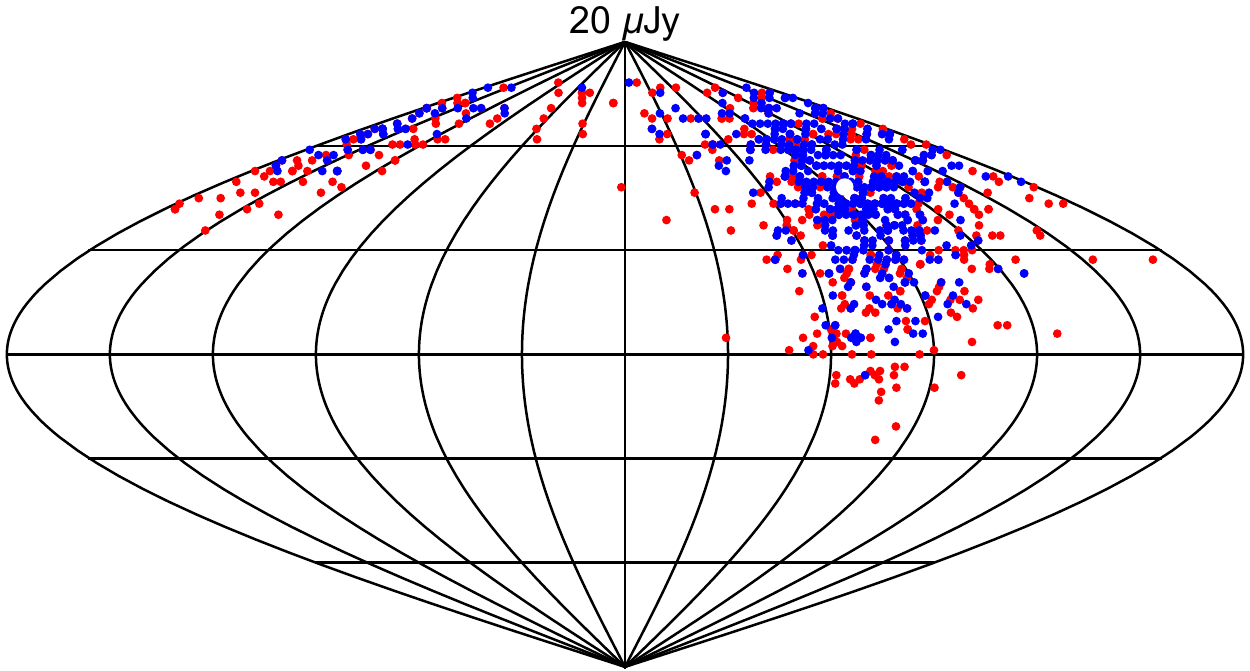}
\caption{Same as Fig.~\ref{fig:dipole_directions_TS}, but using the alternative estimator \eqref{eq:delta-map}.}
\label{fig:dipole_directions_CB}
\end{figure*}

For comparison, we evaluate the constraints using an alternative hemispherical comparison estimator, following~\cite{Bengaly2018b} (see also~\citealt{Bengaly2017, Bengaly2018a}):
\begin{eqnarray}\label{eq:delta-map}
\Delta(\theta) \equiv \frac{\sigma_i^U(\theta) - \sigma_i^D(\theta)}{\sigma} = A \cos\theta \,,
\end{eqnarray}
where 
\begin{eqnarray}
\sigma_i^J = \frac{N^J_i}{2\pi (f_{\rm sky})_i^J}\,, ~~\sigma = \frac{N_{\rm total}}{4\pi f_{\rm sky}}\,.
\end{eqnarray}
Here $i=1,\cdots, 12288$ labels the hemisphere decomposition, corresponding to $N_{\rm side}=32$ grid resolution\footnote{Note that $N_{\rm side}=16$ was chosen in~\cite{Bengaly2017, Bengaly2018a, Bengaly2018b}.}, and $J=U,D$ identifies the `up' and `down' hemispheres in this pixelisation scheme. 

For each hemisphere,  $N^J_i$ is the number of sources, $(f_{\rm sky})^J_i$ is the observed sky fraction, and $\sigma_i^J$ is the source density, with $N^U_i + N^D_i = N_{\rm total}$. $\theta$ is  the angle between the $i$-pixel centre and the observer's motion, so that the maximum $\Delta(\theta)$ value provides the kinematic dipole amplitude, and the pixel centre position where it occurs is regarded as the dipole direction.  

The results are shown in {Fig.~\ref{fig:dipole_directions_CB} and} Table~\ref{tab:dipole_estimates_CB}. It is apparent that the recovered dipole amplitudes are larger than the predicted value \eqref{akin}, and significantly larger than the values recovered via the quadratic estimator \eqref{dip_est_dom}. This  shows that the linear estimator \eqref{eq:delta-map} is 
biased (see \citealt{Rubart2013} for a discussion of different linear estimators that show similar effects).

Despite the bias, this estimator is clearly detecting the signal and recovers the fiducial direction, however with a larger 
spread of values, as is clear from {Fig.~\ref{fig:dipole_directions_CB}}. The advantage of linear estimators is that they are less compute-intense and easy to implement. We thus can 
use them to obtain fast order-of-magnitude estimates of the individual contributions to the cosmic radio dipole.

\begin{table}
	\centering
	\caption{As in Table~\ref{tab:dipole_estimates_TS}, but using the alternative estimator \eqref{eq:delta-map}.}
		\begin{tabular}{ccccc}
		\hline\hline
		Sample & ${S >}$ & $l$ & $b$ & $A$ \\
		& ($\mu$Jy) & (deg) & (deg) & $(10^{-3})$ \\\hline\hline
		full & $1.0$ & $263.16 \pm 20.91$ & $48.12 \pm 11.92$ & $5.46 \pm 0.36$\\
		$z\geq0.5$& $1.0$ & $262.23 \pm 18.69$ & $48.01 \pm 11.05$ & $5.42 \pm 0.31$\\\hline
		full & $5.0$ & $263.89 \pm 24.64$ & $47.65 \pm 14.09$ & $5.56 \pm 0.47$\\
		$z\geq0.5$& $5.0$ & $257.49 \pm 19.41$ & $48.34 \pm 11.40$ & $5.45 \pm 0.33$\\\hline
		full & $10.0$ & $254.05 \pm 40.31$ & $44.96 \pm 17.97$ & $5.64 \pm 0.57$\\
		$z\geq0.5$& $10.0$ & $257.04 \pm 31.51$ & $46.72 \pm 15.91$ & $5.46 \pm 0.38$\\\hline
		full & $20.0$ & $248.83 \pm 55.70$ & $42.91 \pm 21.91$ & $6.14 \pm 1.01$\\
		$z\geq0.5$& $20.0$ & $255.71 \pm 37.45$ & $48.22 \pm 16.73$ & $5.63 \pm 0.46$\\\hline
	\end{tabular}
\label{tab:dipole_estimates_CB}
\end{table}

\subsection*{{Signal to Noise estimate}}

We can make an estimate of the signal-to-noise ratio (SNR) 
for detection of the kinematic dipole, {using the expression in~\cite{Bengaly2018b} (see also~\citealt{Itoh2010,Baleisis1998})}:
\begin{eqnarray}
\label{eq:SNR}
\mathrm{SNR} = \frac{A_{\rm kin}}{\sqrt{A_{\rm LSS}^2 + A_{\rm PN}^2}} \,.
\end{eqnarray}
Here $A_{\rm kin}$ is  given by \eqref{akin},
$A_{\rm LSS}$ is the dipole contribution from large-scale structure, and $A_{\rm PN}$ is the dipole contribution from Poisson noise. 

The last two quantities are estimated using~\eqref{eq:delta-map} on {500} catalogues, constructed as follows:
\begin{eqnarray*}
A_{\rm LSS}: && \mbox{LSS-only simulations, no PN,}\\
A_{\rm PN}: && \mbox{PN-only simulations, no LSS,}
\end{eqnarray*}
where the PN-only maps are homogeneous (no clustering).
The median values of the dipole amplitudes are regarded as $A_{\rm LSS}$ and $A_{\rm PN}$, respectively. We perform this procedure for the full sample, as well as the sub-sample with $z<0.5$ sources excised. This gives a rough estimate of how well the kinematic signal can be detected with the SKA survey specifications. 

The results are presented in Table~\ref{tab:SNR}. It is evident that $A_{\rm LSS}$ is nearly one order of magnitude lower than the kinematic signal for all cases but the full $S>20 \; \mu\mathrm{Jy}$ sample. We also note it becomes larger at higher flux thresholds, albeit the power suppression due to the local structure removal works more efficiently at these cases. Although the shot noise contribution ${A_{\rm PN}}$ slightly increases after removal of $z<0.5$ sources, there is still a significant gain in SNR.


Note that the SKA SNR is much larger than that of the existing surveys NVSS and TGSS, for which $\mathrm{SNR} \simeq 1$~\citep{Bengaly2018b}. 

\begin{table}
\centering
\caption{Clustering and shot noise dipole amplitudes, and SNR \eqref{eq:SNR}.}
\begin{tabular}{ccccc}
\hline
\hline
{Sample} & ${S>}$ & $A_{\rm LSS}$ & $A_{\rm PN}$ & $\mathrm{SNR}$ \\
		& $(\mu\mathrm{Jy})$ &  $(10^{-3})$ & $(10^{-3})$ &  \\\hline\hline
		full & $1.0$ & $0.467$ & $0.049$ & $9.849$ \\
		$z\geq0.5$ & $1.0$ & $0.399$ & $0.052$ & $11.480$ \\\hline
		full & $5.0$ & $0.620$ & $0.089$ & $7.372$\\
		$z\geq0.5$ & $5.0$ & $0.40$ & $0.096$ & $11.197$\\\hline
		full & $10.0$ & $0.662$ & $0.119$ & $6.870$ \\
		$z\geq0.5$ & $10.0$ & $0.417$ & $0.127$ & $10.602$\\\hline		
		full & $20.0$ & $1.347$ & $0.161$ & $3.405$\\
		$z\geq0.5$ & $20.0$ & $0.493$ & $0.177$ & $8.820$ \\\hline
\end{tabular}
\label{tab:SNR}
\end{table}


\bsp	
\label{lastpage}

\end{document}